# Preparation of a clay based superabsorbent polymer composite of copolymer poly(acrylate-co-acrylamide) with bentonite via microwave radiation

Hussam-Aldeen Kalaleh , Mohammad Tally and Yomen Atassi[*]
Laboratory of Materials sciences, Department of Applied Physics, Higher Institute for Applied Sciences and Technology, Po. Box 31983, Damascus, Syria
[*] Corresponding Author: yomen.atassi@hiast.edu.sy

**Abstract:**

In this paper we present a fast, new and easy method for the preparation of bentonite-g-poly(acrylate-co-acrylamide) composite material using microwave radiation. The composite has water absorptivity of 1002 g/g while the corresponding copolymer is tacky and has poor gel texture. The relative thermal stability of the composite in comparison with the copolymer was proved via thermogravimetric analysis These results indicate the synergistic effect of the bentonite with the copolymer by increasing water absorptivity and enhancing thermal and mechanical properties of the final gel. The composite was characterized by X-ray and FTIR. The influence of the environmental parameters on water absorptivity such as the pH and the ionic force was investigated

**Introduction**:

Preparing clay based superabsorbent polymer composite is gaining importance day after day due to enhanced properties and lowering cost of the resulting composite compared with the cost of pure polymer [1]. These composites find applications in horticulture as soil amelioration material in agriculture and in oil recovery processes. Different studies have been conducted to prepare these composites using standard solution polymerization [2], photopolymerization or electron beam methods [3,4].

On the other hand, Microwave irradiation using the commercial household microwave oven has received increasing interest in organic synthesis due to the remarkable enhancement of the rates of some organic reactions over conventional reaction [5]. Microwave energy can be directly and uniformly absorbed throughout the entire volume of the reactive medium, causing it to heat up evenly and rapidly [6]. The microwave irradiation has been successfully used in the graft copolymerization of chitosan with acrylic acid [7].

Bentonite, which is a sedimentary rock consisting to a large proportion of expandable clay mineral called montmorillonite with three-layer structure (smectite), was used extensively in preparing polymer-layered silicate nanocomposites [8]. In this work bentonite was used to prepare cost effective gel through the incorporation of this low-cost raw material, using a domestic microwave oven. The method of preparation is new and easy. The thermal stability of the formed gel was invstigated by thermogravimetric analysis. The grafting of the copolymer onto the montmorillonite was studied using FTIR spectroscopy. The exfoliation of the montmorillonite was demonstrated by X-ray diffractometry. The composite material was tested for the effect of water salinity and environmental pH on water absorbency.

## Experimental

**Materials**

Acrylic Acid (AA) for synthesis (Merck), and Acrylamide (AM) for synthesis (Merck), were used as purchased. Potassium peroxodisulfate $K_2S_2O_8$ (KPS) GR for analysis as an initiator and N,N-methylene bisacrylamide (MBA) for electrophoresis as a crosslinker were also obtained from Merck. Sodium hydroxide NaOH microgranular pure (POCH) was used as a neutralizing agent. Solvents: methanol and ethanol (GR for analysis) were obtained from Merck. Saline sodium chloride (NaCl) (Merck), calcium chloride ($CaCl_2$)(Panreac) and aluminium Chloride ($AlCl_3$) (Merck) were prepared with distilled water. Domestic bentonite from Aleppo was used without further purification. It was only milled to have powder size less than 355 micron, and dried at $120^oC$ until there is no weight loss.

**Instrumental Analysis**

A computer interface X-ray powder diffractometer (Philips, X'pert) with Cu K$\alpha$ radiation ($\lambda$=0.1542 nm) was used to identify the crystalline phases. The data collection was over the 2-theta range of 2 to $20^0$ in steps of $0.02^0$/s. The $d_{001}$ spacing was calculated according to the Bragg's equation: $n\lambda=2d\sin\theta$.

The IR spectra in the 400-4000 $cm^{-1}$ range were recorded at room temperature on the infrared spectrophotometer (Bruker, Vector 22). For recording IR spectra, powders were mixed with KBr in the ratio 1:250 by weight to ensure uniform dispersion in the KBr pellet. The mixed powders were then pressed in a cylindrical die to obtain clean discs of approximately 1 mm thickness.

Thermogravimetric analysis of the bentonite, copolymer and the composite were performed (SETARAM, Labsys TG, TG-DSC $1600^0C$) from room temperature to $600^0C$ at a heating rate of 10 $^0C$/min and under Argon atmosphere.

**Composite Preparation:**

The general procedure for the preparation of superabsorbent polymer composite, through graft copolymerization of poly(AA-co-AM) onto montmorillonite in the presence of MBA and KPS, was conducted as follows: Acrylic acid (6g) was partially neutralized to 80 wt% by addition of NaOH solution (2M) to the acid in an ice bath to avoid polymerization. The solution is then added to Acrylamide solution (6g in ca. 12mL of distilled water). To the above mixture, we added (4g) of bentonite with 0.08g of KPS (in ca. 7 mL of distilled water) and 0.06g of MBA (in ca. 5 mL of distilled water). The total mass of the reactive mixture was brought to 100g by adding distilled water. The mixture is stirred at a stirring velocity of 1500 rpm until it becomes viscous and consistent. This ensures the intercalation of co-monomers in the gallery space of the montmorillonite. Then the mixture is treated in a microwave oven at the power of 950W. The mixture temperature and viscosity increase fast. The gelation point is reached after 90s. The product (as an elastic brown gel) is scissored to small pieces. Then it is washed several times with methanol to dissolve non reacting reagents, and last washed by ethanol, and dried for several hours at $70^0C$ until it becomes solid and brittle. At this point the solid is ground and dried in the furnace at $70^0C$ for 24 hours.

**Copolymer Preparation:**

The procedure for the synthesis of the copolymer poly(AA-co-AM) alone is identical to that of the composite, but without adding bentonite to the reactive mixture.

**Swelling Measurements**

Water Absorbency of the composite is measured by the free swelling method and is calculated in grams of water per one gram of the composite [9]. Thus, an accurately weighed quantity of the polymer under investigation (0.1 g) is immersed in 500mL of distilled water at room temperature for at least 3 hours. Then the swollen sample is filtered through weighed 100-mesh (150 μm) sieve until water ceased to drop. The weight of the composite containing absorbed water is measured after draining for 1 hour, and the water absorbency is calculated according to the following equation: [10]

$$Q = (m_2-m_1)/(m_1-m_0)$$

Where $m_0$, $m_1$ and $m_2$ are respectively the weights (g) of the clay in the sample, the dry sample and the swollen sample.

**Results and Discussions**:

**FTIR spectra**: Figure 1 shows the FTIR spectra of the bentonite, the copolymer, and the composite material.

To understand the formation and crosslinking of hydrogel obtained from microwave irradiation of the mixture of AA, AM co-monomers and the bentonite, we will compare the FTIR spectra of the bentonite, the copolymer, and the composite material.

In the spectrum of the bentonite, the peaks observed were at 3614 $cm^{-1}$ and 3547 $cm^{-1}$ for the OH groups stretching in Al-OH and $Fe^{3+}$-OH respectively, and at 1029 $cm^{-1}$ corresponding to Si-O stretching. The peaks at 1655 $cm^{-1}$ and at 3414 $cm^{-1}$ are assigned to OH deformation and OH stretching in H-O-H respectively.

In the spectrum of the copolymer hydrogel, the peaks observed at 3443 $cm^{-1}$ and at 3184 $cm^{-1}$ correspond to O-H and N-H stretching respectively. The absorbance at 2925 $cm^{-1}$ is assigned to –C-H stretching of the acrylate group. The peak at 1676 $cm^{-1}$ and at 1726 $cm^{-1}$ are assigned to C=O stretching of the acrylamid groups and acrylate groups, respectively. The peak at 1168 $cm^{-1}$ is attributed to –COO stretching of the acrylate group and the peak at 1559 $cm^{-1}$ is assigned to –COONa group.

When the mixture co-monomers and bentonite was irradiated to produce a gel, the corresponding FTIR has the same peaks corresponding to AA, AM and bentonite, but with deviation to smaller wavenumbers.

It is suggested that the montmorillonite crosslinked through hydrogen bonding and entrapped in the poly(AA-co-AAm) hydrogel with IPN network [11].

Furthermore, if we examine the spectrum of the composite material, we find the disappearance of the peaks corresponding to the OH groups stretching in Al-OH and $Fe^{3+}$-OH in the montmorillonite structure, while these peaks appear clearly in the spectrum of the blend (prepared by physical mixing of the copolymer with the bentonite with the same proportion as in the composite material), Figure 2. This disappearance may be contributed to the esterification between the acrylic acid and the free –OH groups at the montmorillonite. The band at 1730 $cm^{-1}$ corresponding to C=O stretching of the ester group is partially overlapped by the band at 1726 $cm^{-1}$.

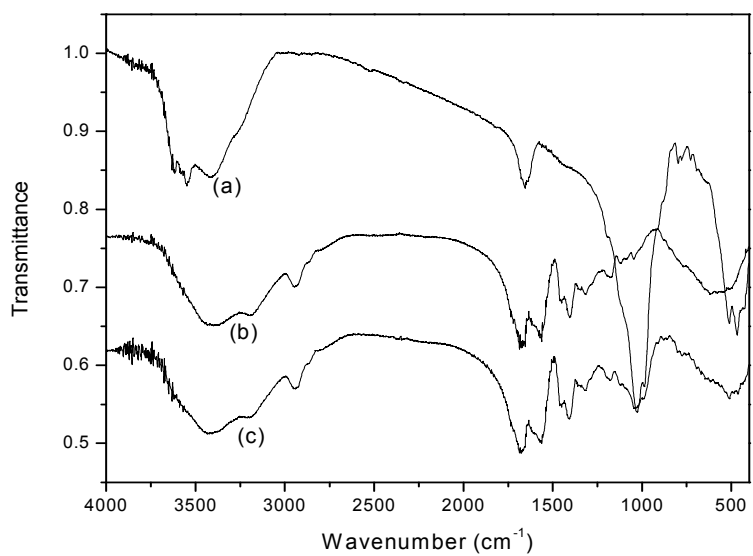

Figure 1: FTIR spectra of (a) Bentonite, (b) Copolymer and (c) Composite material.

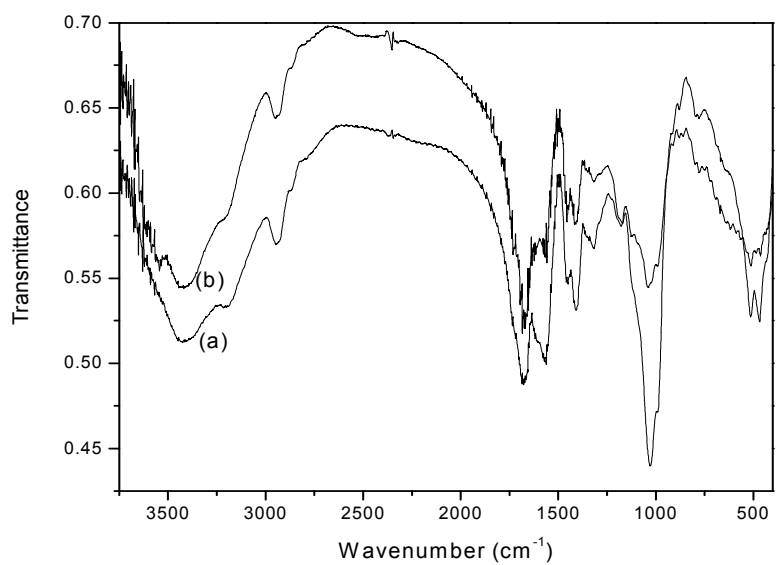

Figure 2: FTIR spectra of (a) Composite material and (b) blend: physical mixture of the copolymer and bentonite with the same proportion as in the composite material.

**X-Ray diffractograms**: Figure 3 shows the XRD patterns of the bentonite and the composite material in the range $2^0$-$20^0$.
The original bentonite exhibits a peak associated with a spacing of 1.497 nm at $5.9^o$, whereas, the absence of this basal peak suggests a high dispersion of clay platelets (exfoliation) in the composite material. In the composite we assume that the gallery structure of the montmorillonite breaks due to pressure exerted by the intercalated co-monomers, leaving behind upon co-polymerization an exfoliated composite material.

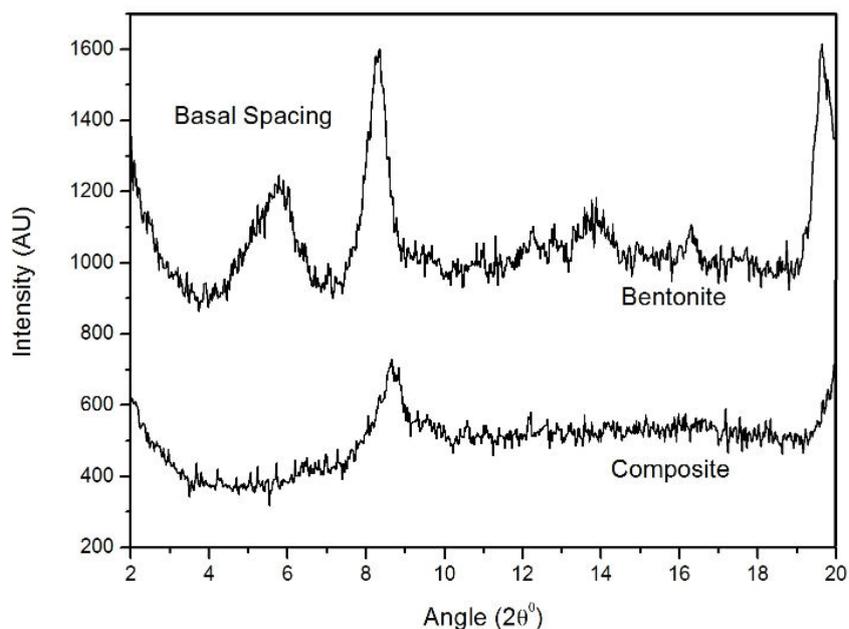

Figure 3, XRD diffraction patterns of the bentonite and the composite material, reports the spacing between ordered layers of clay via the presence of the $d_{001}$ or basal spacing and absence of this basal spacing in the composite material.

By combining the results of FTIR spectroscopy and XRD patterns, we can conclude that the grafting of the co-monomers on the montmorillonite was carried out in aqueous solution.

The co-monomers can be adsorbed onto the surface of the montmorillonite galleries in two ways: first, by formation of hydrogen bonds between their amide and acid groups and water molecules surrounding the exchangeable cations and, second, by formation of ion-dipole interactions between the acrylate groups and the interlayer exchangeable cations. These interactions lead to an increase in the basal spacing relative to montmorillonite in bentonite and eventually lead to exfoliation [12]. Experimentally, these interactions between the commoners and the bentonite leading to exfoliation manifested experimentally by thickening of the solution after vigorous stirring.

On the other hand, co-monomers can be adsorbed onto the clay by either formation of hydrogen bonding between the amide and acrylic acid groups and the hydroxyl groups on the edges of montmorillonite platelets or by esterification. These interactions have no effect on the basal spacing.

**Thermal stability:**

Figure 4 illustrates the thermogravimetric curves (TGA) of the bentonite, the copolymer alone and the composite material.

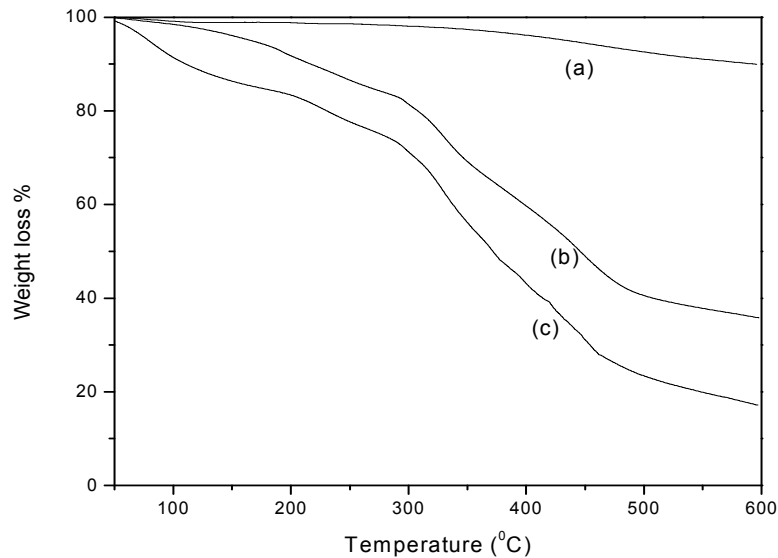

Figure 4: Thermogravimetric curves of (a) Bentonite, (b) Composite material and (c) Copolymer.

The weight loss due to the formation of volatile products after degradation at high temperature is monitored as a function of temperature. As the heating occurs under an inert gas flow, a non-oxidative degradation occurs. Figure 4 shows clearly that the grafting of the copolymer on the montmorillonite enhance the thermal stability by acting as an insulator and mass transport barrier to the volatile products generated during decomposition. The onset temperature of the degradation is about $50^0C$ higher for the composite and it remains less susceptible to degradation than its corresponding copolymer along the thermal profile.

**Water absorptivity:**
Water absorptivity of the composite was measured by the free swelling method, and it was 1002 g/g, while the corresponding copolymer was tacky and has poor gel texture. Other measurements were carried out on a similar composite, prepared with 60% of neutralization of the acrylic acid and 0.12 g of MBA, showed a water absorptivity of 606 g/g while the absorption of the corresponding copolymer was only 452 g/g. These results indicate the synergistic effect of the bentonite with the copolymer in increasing water absorptivity. The exfoliation of the montmorillonite could be a benefic factor in increasing the specific surface, and therefore increasing water absorptivity.

**Effect of the environmental parameters on water absorptivity**
**-Effects of salt solution on water absorptivity:**
The composite was tested for the effect of water salinity on its swelling capacity. Different concentration of NaCl, $CaCl_2$, $AlCl_3$ solutions were prepared in order to study the effect of ion charge and ion concentration on water absorption [13].

| Water Absorptivity g/g | 0.9%w | 0.09%w | 0.009%w |
|---|---|---|---|
| NaCl | 77 | 200 | 516 |
| $CaCl_2$ | 30 | 84 | 326 |
| $AlCl_3$ | 6 | 14 | 114 |

The table above shows that water absorption decreases with increasing the ionic strength of the saline solution as cited in Flory equation [14]. The ionic strength of the solution depends on both the concentration and the charge of each individual ion. In fact the presence of ions in the solution decreases the osmotic pressure difference, the driving force for swelling, between the gel and the solution. In addition, multivalent cations ($Ca^{2+}$ and $Al^{3+}$) can neutralize several charges inside the gel by complex formation with carboxamide or carboxylate groups, leading to increased ionic crosslinking degree and consequently loss of swelling. The effect of cation charge and its concentration on swelling can be concluded from Figure 5.

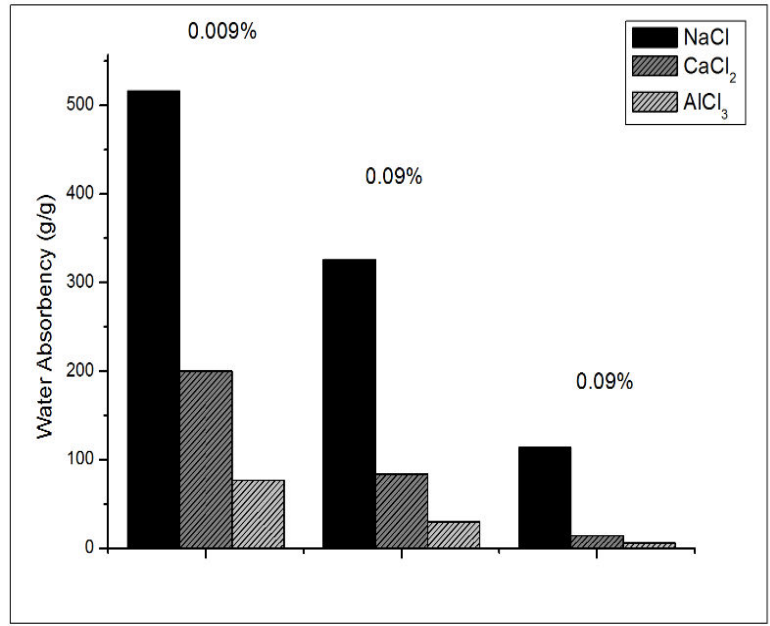

Figure 5: Effect of salt solution on water absorptivity.

**-Effects of the environmental pH on water absorptivity:**

Studies have indicated that water absorption of hydrogels are sensitive to environmental pH [15]. So, the swelling behavior of the composite material was studied at various pH values between 2.0 and 13.0, at room temperature, Figure 6. To clearly observe the net effect of pH, buffer solutions containing lots of ionic species, were not used as swelling media, as the swelling is strongly decreased by ionic strength. Therefore a stock of concentrated solution HCl and NaOH were diluted with distilled water to reach the desired acidic or basic pH.

At low pH, the swelling capacity decreases as the sodium carboxylate group on the polymer network is protonated. The polymer shrinks and becomes hydrophobic. This in turn decreases the degree of ionization and hence decreases the swelling ratio.

At high pH, the swelling capacity also decreases by "charge screening effect" of excess $Na^+$ in the swelling media, which shields the carboxylate anions and prevents effective anion-anion repulsion.

In the interval pH=4 to pH=8, some of carboxylic acid groups are ionized and the electrostatic repulsion between $COO^-$ groups causes an enhancement of the swelling capacity.

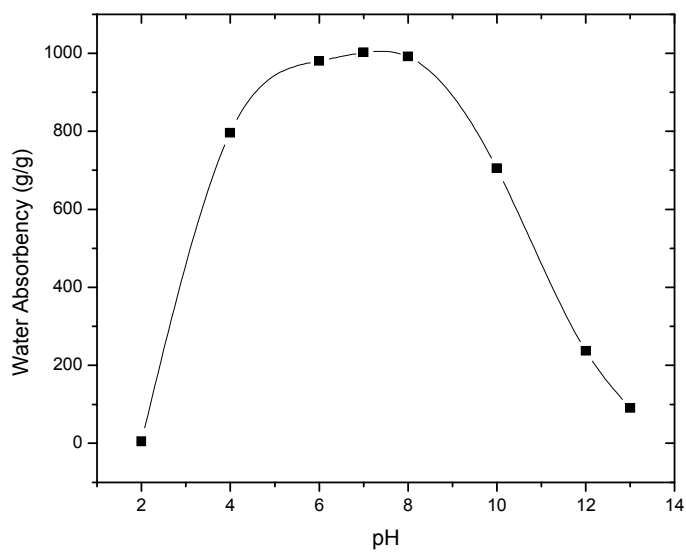

Figure 6: Effect of environmental pH on water absorptivity.

**Conclusion**

Bentonite based superabsorbent polymer composite was synthesized using a domestic microwave oven. The incorporation of this inexpensive clay within the polymeric matrix enhanced the thermal stability of the superabsorbent polymer composite and its water absorptivity. The FTIR spectrum of the composite material illustrated both the peaks characteristic of co-monomers and the bentonite. The grafting was attributed to hydrogen bonding and esterification. XRD pattern of the composite confirmed the exfoliation of the clay by the disappearance of the peak at $5.9°$ with interlayer spacing d=1.497nm. The increase of the composite material water absorption in comparison with its corresponding copolymer could be attributed to the exfoliation of the montmorillonite, and therefore augmentation of the specific surface of the final product.